\documentclass[manuscript]{aastex}

\newcommand{\beq}{\begin{equation}}
\newcommand{\eeq}{\end{equation}}

\newcommand{\kms}{\mbox{ km s$^{-1}$}~}
\newcommand{\cms}{\mbox{ cm s$^{-1}$}~}

\newcommand{\gcmss}{\mbox{g cm$^{2}$ s$^{-1}$}~}

\newcommand{\Mo}{\mbox{M$_{\odot}$}~}

\newcommand{\Ro}{\mbox{R$_{\odot}$}}

\shorttitle{Single Stars cannot do it}
\shortauthors{Garc\'{\i}a-Segura et al.}

\begin{document}


\title{Single Rotating Stars and the Formation of Bipolar Planetary Nebula}


\author{G. Garc\'{\i}a-Segura}
\affil{Instituto de Astronom\'{\i}a, Universidad Nacional Aut\'onoma 
de Mexico, Km. 103 Carr. Tijuana-Ensenada, 22860, Ensenada, B. C., Mexico}
\email{ggs@astrosen.unam.mx}

\author{E. Villaver}
\affil{Departamento de F\'{\i}sica Te\'orica, Universidad Aut\'onoma de Madrid,
       Cantoblanco, E-28049 Madrid, Spain}

\author{N. Langer }
\affil{Argelander-Institut f\"ur Astronomie, Universit\"at Bonn, D-53121 Bonn, Germany}

\author{S.-C. Yoon}
\affil{Astronomy Program, Department of Physics and Astronomy,
Seoul National University, Seoul, 151-747, Republic of Korea}

\and

\author{A. Manchado\altaffilmark{1,2}}
\affil{Instituto de Astrof\'{\i}sica de Canarias,  Via L\'actea s/n, E-38200 La Laguna, Tenerife, Spain}

\altaffiltext{1}{Departamento de Astrof\'{\i}sica, Universidad de La Laguna,
E-38206 La Laguna, Tenerife, Spain}
\altaffiltext{2}{Consejo Superior de Investigaciones Cient\'{\i}ficas (CSIC), Spain}



\begin{abstract}

We have computed new
stellar evolution  models that include the 
effects of rotation and magnetic torques under different
hypothesis. The goal is to
test if a single star can sustain in the envelope the 
rotational velocities needed for the magneto hydrodynamical (MHD)
simulations to shape bipolar
Planetary Nebulae (PNe) when the high mass-loss rates take place. Stellar
evolution models with main sequence masses of 2.5 and
5 \Mo, and initial rotational velocities of 250 \kms have been
followed all the way
to the PNe formation phase. We find that stellar cores have to be spun
down using magnetic torques in order to reproduce
the rotation rates observed for white dwarfs. During the asymptotic
giant branch phase and beyond, the magnetic braking of the core
has a practically null effect in increasing the rotational velocity of
the envelope since the stellar
angular momentum is removed efficiently by the wind. 
We have, as well, tested best possible case
scenarios in rather non-physical contexts to give enough angular
momentum to the envelope. We find that  we
cannot get the envelope of a single star rotating at the speeds needed by the 
MHD simulations to form bipolar PNe.  We conclude that single stellar
rotators are unlikely to be
the progenitors of bipolar PNe under the current
MHD model paradigm.

\end{abstract}



\section{INTRODUCTION}

The overall details of the Planetary Nebulae (PNe) formation process are
well understood since  Kwok, Purton \& Fitzgerald (1978)
proposed that
it takes place due to the collision of two winds: a slow one
characteristic of the red giant phase and a faster one which is
produced later, by the naked stellar nucleus (see also
Kahn \& West 1985; Balick 1987). Within this framework,
PNe modeling efforts have mostly focused on the shaping process
itself, since understanding the
formation of the complex PNe morphologies remains one of the open
questions in the field (see reviews by Pottasch 1984; Iben 1993;
Balick \& Frank 2002; Shaw 2012). 

Models have achieved a good deal of sophistication in reproducing
morphological features using  the hydrodynamic collimation of the fast wind
by an equatorial density enhancement mechanism first implemented by Icke (1988) and
Icke et al. (1989). In these models, the
density enhancement responsible for the collimation is built
ad-hoc, as it is in the Mellema, Eulderink,
\& Icke (1991), Icke, Balick, \& Frank (1992), Frank \& Mellema
(1994), and 
Dwarkadas, Chevalier, \& Blondin (1996) models. 

Nowadays that is clear that the aspherical slow-fast wind combination can reproduce most PNe
morphologies, the problem is to understand how, or if, a single star
can create the
equatorial density asymmetries during the Asymptotic Giant Branch
(AGB) phase. The greatest challenge is to generate asymmetric stellar outflows via
the combination of the winds when they are constrained by stellar
evolution.

The first attempt to link the density enhancement to the evolution of
the star was made by Garc\'{\i}a-Segura et al. (1999) by using a
rotating star with (or without) a magnetized wind and the wind
solution by Bjorkman and Cassinelli (1993). Matt et al. (2000) have
also shown how an isolated AGB star can produce a dense equatorial
disk using a dipole magnetic field on the surface of the star.   A rotating
star also provides a natural solution to the problem of creating an
AGB density enhancement with different degrees of collimation
(Garc\'{\i}a-Segura et al. 1999).  Even the magnetic field topology is
derived from what is expected in the solar wind, where the poloidal
field lines connecting the outflowing wind to the solar surface are
wound up due to the rotation of the Sun.  So, it seems that, so far,
most scenarios for single stars require the key ingredient of stellar
rotation either as a starting point for the solution, as a consequence
of the evolution, or just as a tool to generate the magnetic field.

The literature is abundant in calculations of massive star
evolution with rotation and rotationally-induced processes (see e.g. Maeder  \&
Meynet 2000; Heger et al. 2000) but very few computations for low-mass
stars through the AGB phase, including the effect of rotation are  available. 
Stellar evolution calculations
for 1 to 3 \Mo stars with rotation have been computed to the early AGB
phase by Suijs et al. (2008) to deduce the white dwarf (WD)
angular momentum distribution.  Recently, Prinja et al. (2012) have also presented
the evolution of a 1.5 \Mo rotating star in their study of the PN  NGC
6543. Both works arrived at the same general conclusion: that magnetic
torques are needed to transfer the angular momentum from the
rapidly spinning cores to the envelope in order to reproduced the
observed rotational velocities of WDs (Kawaler 2004). 
In this paper, we focus on studying if the mechanism capable of reducing the spin
of the core to match empirical values  (a magnetic torque) is able to spin up
the envelope to the values required by the MHD calculations. 

We present new calculations of the evolution of  2.5 and
5 \Mo  stars from the Zero-Age Main Sequence (ZAMS) to the post-AGB phase with rotation and
magnetic torques using  a realistic prescription for the mass-loss
rate during the AGB phase. 
Motivated by  the analytical estimates by Garc\'{\i}a-Segura et al. (1999), the
goal is to study whether we can obtain the 
rotation at the stellar surface needed to shape
bipolar PN, by transferring under
different hypothesis, the angular momentum from the core. 

The paper is organized as follows: the stellar evolutionary code
is described in \S 2; the results of the calculations with and without
magnetic torques are presented in \S 3 where we also provide the
description of the models where the magnetic torques are switched on
at selected times;  a full discussion and the conclusions of the paper are provided in  \S
4. 

\section{STELLAR MODELS}
The stellar evolution calculations have been done using the Binary
Evolution Code (BEC) (Petrovic et al. 2005; Yoon et al. 2006).
BEC is a one dimensional hydrodynamic, stellar evolution code 
designed to evolve stellar models of single, and binary stars, which descends from the code
for binary stars (Braun \& Langer 1995) and from the code STERN (Langer 1991). 
The code includes diffusive mixing due to convection, semi-convection (Langer et al. 1985), and 
thermohaline mixing as in Wellstein et al. (2001). This code includes the effect of the centrifugal 
force on the stellar structure, and time-dependent chemical mixing and transport of angular 
momentum due to rotationally-induced instabilities (Heger et al. 2000). 
We also include chemical 
mixing and transport of angular momentum due to magnetic fields (Spruit 2002), as in 
Heger et al. (2005) and Petrovic et al. (2005). 
The rotation profile in the convection zone is self-consistently calculated by solving 
the diffusion equation within the code.

We calculate the evolution of stars with initial stellar masses of 2.5
and 5\Mo and solar metalicity. Two main reasons motivate the choice of
the initial progenitor masses. 
First, the single star hypothesis
for the formation of bipolar PNe has been suggesting higher mass
stellar progenitor for this morphological class as indicated by
their chemical abundances (see e.g. Stanghellini et al. 2006; Manchado
et al. 2000), and their closer distribution to the galactic plane 
(Corradi \& Schwarz 1995). Second, we aim
to constrain the mechanism for the most favorable scenario and
faster rotators are found among the main
sequence (MS) stars with the higher masses within the PNe progenitor mass range. 

We have adopted an initial, representative equatorial rotation velocity of 250 \kms
at the ZAMS (Fukuda 1982). The mass-loss rate
used is the one by Reimers (1975) for the Red
Giant phase with an $\eta = 0.5$ parameter and that by Vassiliadis \&  Wood (1993) for the AGB
phase. For the 2.5 \Mo star we have computed the evolution from
the ZAMS to the post-AGB phase to the point were the remaining stellar mass
is 0.67 \Mo. This star loses 1.83 \Mo through the mass-loss at the AGB
thermally pulsating phase. The 5\Mo model calculation is stopped after 19 thermal pulses beyond which the
CO core angular momentum does not change significantly (see
also Suijs et al. 2008).

\section{RESULTS}

The evolution of a ZAMS 2.5 \Mo star in the Hertzsprung-Russell
(HR)-diagram is shown in Figure 1.  The track shows how
the star undergoes the major evolutionary phases, MS, Red Giant Branch (RGB) and  AGB. It
also shows the characteristic modulations associated to
the thermal-pulses taking place in the stellar interior at
the end of the AGB phase. As mentioned
before, the mass-loss rate mode has been changed between the RGB and AGB
phases from the Reimers prescription to the Vassiliadis \& Wood (1993)
parameterization which has a dependency with the pulsational period of
the star which in turn translates into a period-mass-radius relation.
The stellar evolution is followed well into the
post-AGB phase (for the meaning of the marks in the plot see \S 3.1).
The model in Figure 1 has been computed using magnetic torques
and with an initial rotational velocity of 250 \kms. The evolution of
the model without magnetic torques is very
similar. The only significative difference is that the modulations
associated to the thermal pulses at the tip of the AGB occur at slightly different locations
in the HR diagram. 

The evolution of the surface equatorial rotational velocity is
presented in Figure 2  for the 2.5 \Mo star and the 
non-magnetic case.  The top panels show the evolution up to
the early thermal pulsing (TP) AGB phase and the bottom panels 
%usas TP AGB antes...definelo antes y usa TP-AGB
the evolution of the rotational velocity until the end of the
calculation when the star is well into the post-AGB phase. Note the difference in
scale between the top and the bottom panels in the figures.

Along the MS evolution of the star, the density increases
in the stellar core leading to a more efficient nuclear burning, and to
a tiny increase in the stellar radius. This causes a decrease in
the rotational velocity along in the first $\approx 5\times 10^8$ \,yr of the
evolution (the largest radius reached at the MS is 3.8 \Ro) as seen in the top panel of
Figure 2. 

The point of H-core burning exhaustion induces a small peak in the
surface rotational velocity which is immediately followed by an abrupt
drop consequence of the expansion in the radius along the RGB (up to 25
\Ro). During the He-core burning phase the surface rotational velocity
remain at a value of 15 \kms with the stellar radius during this
phase remaining at a nominal value of 8.4 \Ro \, . 
The evolution during the AGB phase
begins past  $7.4 \times 10^8$ \,yr and it is characterized by stellar
expansion, heavy mass-loss rates, and thermal-pulses. The multiple peaks
in the bottom of Figure 2 are associated to the increase in radius
experienced by the star in the aftermath of the pulses. The surface
rotational velocity at the last point shown at the bottom of Figure 2 
(that of a post-AGB star with 6400 \,K)  is almost zero.

We find that when the 2.5 \Mo star, evolving without magnetic field-induced internal angular
momentum transport, reaches the
post-AGB phase, it has lost most of its angular momentum and
its surface rotation velocity is 9.7 $\times 10^{-6}$\kms.  Similar results
are obtained for the 5 \Mo star which reaches a surface rotation
velocity of  $ 10^{-6}$\kms at the end of the AGB phase. 

Figure 3 shows the evolution of a 2.5 \Mo star including the effects of angular momentum transport due to
magnetic fields (see Spruit 2002). The overall behavior
of the surface equatorial rotational
velocity is similar to the non-magnetic case shown before. Small
differences are found for instance in the velocity peak reached at the end of
the MS, or in the smaller velocity in the surface at the
beginning of the AGB, and are due to
the more efficient transport of angular momentum between the core and
the envelope (Heger \& Langer 1998). The slightly different evolutionary timescales between the
magnetic and non-magnetic models are caused by the same effect. 

In this calculation using magnetic torques the
surface rotational velocity  is almost zero,  8.4 $\times
10^{-6}$\kms, at the post-AGB phase when the star has
reached a temperature of 6400 \,K. The magnetic 5 \Mo star reaches a
surface rotation velocity at the end of the computed evolution of  4 $\times
10^{-6}$ \kms . 

In Figure 4 we show snapshots of the distribution of the rotational velocities
within the stellar structure taken at key evolutionary times. The
non-magnetic model is plotted on the left, and the magnetic
model in the right panels. From top
to bottom are shown the ZAMS, the end of the MS, RGB, 
He-core burning phase, and the post-AGB phase when
the star has reached a  temperature of 6400 \,K. The top
panels, taken at the ZAMS, represent the initial conditions for both models, a
constant angular velocity with a solid body rotation of v$_{surf}$ = 250 \kms. 

If we focus in the left panels  showing the non-magnetic model, we see that at the end of H-core
burning, the recently formed He-core has been slightly 
spun up by contraction (angular momentum conservation) to ~40 \kms
while the surface slows down by the expansion of the envelope. Note that the
radiative envelope still rotates as a solid body. By the end of the 
RGB phase (H-shell burning), the third panel from the
top shows that the dense He-core has been spun up to 150 \kms by a much
larger contraction, while the expansion of the envelope 
to a radius of 25 \Ro~has slowed down the surface velocity
to a value of $\approx$ 5 \kms. By the end of the He-core burning,
fourth panel, the core is slowed down by redistribution of 
angular momentum. The core rotates
at ~55 \kms, while the surface is rotating at only ~10 \kms. 

By the time the star reaches the  AGB phase the contraction of the
still forming CO core  produces a spin-up at the center, while the
large expansion of the convective envelope, up to 500 \Ro,  slows down
completely the stellar surface. This results in a star that at the
post-AGB phase has a fast rotating CO degenerate core 
that will finally end up as a very fast (~150 \kms) spinning WD.  

The magnetic model in the right panels of Figure 4  show  the action of
magnetic torques between core and envelope operating along the whole
evolution. This is evident at the end of the RGB phase, where
the star as a whole has completely slow down its rotation to values under 4
\kms. Note instead the core speed of 150 \kms of the model without magnetic torques. At
the post-AGB phase, the core only rotates at 5 \kms. This value of
the rotational velocity match better those obtained through
asterosismological observations of ZZ Ceti stars (Bradley 1998,
2001;Dolez 2006; Handler 2001, Handler et al. 2002; 
Kepler et al. 1995; Kleinmann et al. 1998; Winget et al. 1994). 
As already pointed out by Suijs et al. (2008),
magnetic torques are required to reproduce the slow rotation rates observed
for WDs.

\subsection{Hypothetical Cases}

Let us assume that we have a 2.5 \Mo non-magnetic model up to the AGB phase and that
the angular momentum of the core can be transported to the envelope 
at the end of the evolution. This will be along the lines of  the more
``optimistic'' case discussed in Garc\'{\i}a-Segura et al. (1999) for
the formation of bipolar PNe (see below). 
The CO core has a mass of 0.656 \Mo and it is rotating at 150 \kms. It
is logical then to assume that an 
important change in the surface rotation will occur once the angular momentum is transported 
outwards.  To analyze this hypothetical scheme, we have explored three
cases computed without magnetic braking until the stellar mass reaches
2.45, 1.85, and 0.674 \Mo respectively. At which point the magnetic braking is switched on. 
These models are labelled A, B and C respectively. The location in the HR diagram where the 
magnetic braking is turned on is marked in Figure 1  (see also Figure 5).  Physically these
models correspond to the beginning of the thermal pulses (model A), to
the next to last thermal pulse (model B), and to a few thousand years before the PNe
formation. 

In Figure 5 we show the evolution of the helium burning luminosity as
a function of the stellar mass for the model without magnetic braking. 
It is important to note that more than half of the stellar mass is lost in just
the last two thermal pulses (see also Vassiliadis \& Wood 1993), 
and that there are no meaningful differences regarding the mass-loss
between the magnetic and non-magnetic models.

Figure 6  shows the rotational structure of the star at the post-AGB phase, when the star is in
the blueward evolution and has reached a temperature of 6400 K, just a few thousand years before
the PN becomes ionized.
The CO core of model A rotates at 6 \kms which agrees with the typical
rotation velocities observed in 
WDs. The cores of models B and C  rotate at 17 and  80 \kms respectively, 
values above the observational limit of 10 \kms (Suijs et al. 2008) .  

The late turn on of the magnetic braking used in models A and B 
has no appreciable effect on the rotational velocities
reached on the surface of the star. 
Only for model C we find a noticeable increase of the surface
rotation. However,  model C rotates at only 1.8 \kms by the time the
star reaches the post-AGB phase (at 6400 \,K), 
while the escape velocity has increased up to  $\sim 50$ \kms. A faster rotation than
that reached would be
required in order to produce asymmetries by the wind compressed model 
(Bjorkman \& Cassinelli 93; Ignace et al. 1996).

It is interesting to note in Figure 5 that the Helium production of energy
during the thermal pulses is increased up to 7 orders of magnitude but is never 
turned off completely during the inter-pulse phases. The fact that
energy production in the helium shell is never completely switched off
in the evolution rules out the ``optimistic'' scenario
proposed by Garc\'{\i}a-Segura et al. (1999). The consequence is that an entropy barrier
always remains in the inter-pulse phase in the non-magnetic
model.

An entropy barrier, i.e., a positive entropy gradient, is
formed, e.g., at the location of a nuclear burning shell due to the production
of heat. In a chemically homogeneous situation, this implies a stabilizing
effect (buoyancy) which is proportional to the entropy gradient (Kippenhahn
\& Weigert 1990), implying that the larger the entropy
gradient, the more stable is the stratification. Any radial gas motion
present in this case will be damped,  giving  rise to a dynamically stable
configuration ({\em Schwarzschild criterion}). 
If there is a chemical gradient present in the plasma ($\mu$-barrier) its
effect is added to the effects of the entropy gradient. As a consequence, a negative
$\mu$-gradient, such as the one present in our our calculations (see
also Langer et al. 1999; Heger et al. 2000) of a CO-core surrounded by
an envelope of H and He is highly stabilizing ({\em  Ledoux criterion}).
Note that convective mixing occurs
when the entropy is rather constant, i.e., in adiabatic conditions, or
when lower-entropy gas lies above higher-entropy gas, since the adiabatic
gradient is normally positive.

The ``optimistic case'' involves an scenario in which the He-core evolves decoupled from the 
envelope and retains its
angular momentum, i.e., the entropy barrier of a nuclear burning 
shell prevents that angular momentum can leak out of the core.
When the star move to the TP-AGB phase, the H and He 
burning shells are alternatively switched on and off. Thus the plausible barriers vanish 
periodically and core-envelope angular momentum exchange could occur during this stage.
But, as Figure 5 shows, there is a stable minimum burning configuration
that prevents the total leak to the envelope of the core angular momentum.

The effect of the drain of some angular momentum
out through the H-and He-shells during the TP-AGB is visible in Figure 7 on
the little spikes in the $ j_s(t)$ curve, where is shown  
the specific angular momentum of the surface layers
calculated as $ j_s = R \times v_{surf} $. 
These small leaks are just buried under the stronger effect of the heavy
winds, especially towards the end of the evolution when most of the
mass is carried away from the star.

Figure 8 shows the evolution of the stellar radius and the 
surface equatorial rotational velocity for the model without magnetic
torques. It is shown only the interval from the last three 
thermal pulses to the post-AGB phase when the star reaches a temperature
of 6400 \,K. 
Note that the stellar material that will form the optical ionized nebula
is the material ejected only during the last pulse (see Villaver et al. 2002)
and at this time 
the rotation velocity in the stellar surface is of the order of $\sim 10^2$ \cms.

\section{DISCUSSION AND CONCLUSIONS}

We have conducted a series of stellar evolutionary calculations in order to
test if we can obtain, from a rotating ZAMS star, the velocities
in the stellar surface at the tip of the AGB evolution
that MHD models need to form bipolar PNe. 
Rotational speeds above 80 \% of
the critical rotation value are needed (the closer to critical
rotation the better, where the $\Omega$ limit or break up speed is
achieved; see Garc\'{\i}a-Segura et al. 1999; Ignace et al. 1996). The critical
rotation  is defined as, $v_{\rm crit} = \sqrt{GM(1-\Gamma)/R}$ where, $G$ is the
gravitational constant, $M$ and $R$ the stellar mass and  radius. The term
$\Gamma=L/L_{\rm edd}$  is just the ratio between $L$ the stellar
luminosity and $L_{\rm edd}$ the
Eddington luminosity. For AGB winds, $\Gamma$ should be close to unity by
definition, so, one would expect that the critical velocity should be
close to zero or approaching it. However, since the sound speed at the
surface of an AGB star is of the order of $\sim 1 $ \kms, values below that
for the rotation will not have any impact in the wind compression
mechanism (Bjorkman \& Cassinelli 93; Ignace et al. 1996).
We are obtaining from the stellar evolutionary calculations, 
rotational speeds in the surface at the end
of the AGB of at most $1 \times 10^{-5}$ 
\kms which are too small and are independent on whether
we include or not magnetic field induced angular momentum transport in
the calculations. Note that the value given from model C of 1.8 \kms is for
the post-AGB phase when the star has 6400 \, K. At the TP-AGB phase, model
C also has $1 \times 10^{-5}$ \kms .
It is interesting to mention that the model by Dorfi \& H\"ofner (1996) also
require velocities from 3 to 7 \kms to become efficient. 

Based on our calculations, we can now argue against the formation of bipolar nebulae
by the effects of rotation from single stars.
For that, we  compare the initial angular momentum
with the one obtained at the end of the AGB phase. 
The initial total angular momentum of the 2.5 \Mo star is
$J_{\rm ZAMS}   = 7.095 \times 10^{50} $ \gcmss . 
At the end of the AGB, when $t=7.80637 \times 10^8$  yr (see Figure 8),  the values for 
the radius and rotational velocity are  $ R = 463 $ \Ro \, and
$v_{surf} = 296 $ \cms. 
That time represents the moment the
star reaches the largest 
radius just before the last thermal pulse 
and when the mass-loss rate is the largest for a longer period of  time.
Integrating the stellar structure, we then obtain a total angular momentum 
of  $J_{\rm AGB}    = 5.667 \times 10^{48} $ \gcmss .
The decrease of two orders of magnitude, i.e., a factor of 125, 
is due to mass loss (Heger \& Langer 1998).
This angular momentum can be split in  $J_{\rm core}   = 4.018 \times 10^{48} $ \gcmss and
$ J_{\rm env}    = 1.649 \times 10^{48}  $ \gcmss , for the core and the envelope respectively,
where the CO core has a mass of $M_{\rm core} = 0.656 $ \Mo 
and the envelope  $M_{\rm env} = 0.352 $ \Mo, being the total mass $M_{\rm tot} = 1.008 $ \Mo.

The rotational velocity at the end of the AGB phase is 296 \cms,  and
at least 2 \kms are needed to have a mechanism capable of producing
asymmetries by the AGB slow wind. This implies that we need to achieve
a rotational velocity a factor of 676 larger. 
Assuming rigid rotation because of convection, increasing
the surface velocity by this factor implies to increase in the same factor the 
total angular momentum of the envelope. This gives a new value for
the ``ideal''  envelope angular momentum of $ J_{\rm ideal}  = 1.115 \times 10^{51} $ \gcmss ,
which is a factor of  1.57 larger than the ZAMS angular momentum of the whole star
($J_{\rm ZAMS}$). This is impossible to achieve for a single, isolated star,
since the angular momentum at any given time cannot be larger than the initial one. 
Note also that achieving this
angular momentum for the envelope at the end of the AGB would require
the star at the ZAMS to be rotating at velocities
orders of magnitude larger than the break up limit which for this
stars is of the order of 300  \kms (Collins 1974).

We find that the amount of angular momentum needed to operate a MHD mechanism at
the stellar surface cannot be provided
by the star itself. However, it could be easily injected externally
by a stellar or substellar companion. Just a 6 jupiter mass planet at
an orbital distance of 5 AU has this amount of angular momentum and it
has been shown can be easily engulfed into the stellar surface by
tidal forces reaching far beyond the stellar radius (Villaver \& Livio
2007, 2009; Mustill \& Villaver 2012). The effects on the stellar rotation
of binaries in general will be the subject of a future article.

We find that the inclusion of angular
momentum transport through magnetic fields, and the coupling between
the rapidly rotating core and  the slowly rotating envelope produce
the right amount of core spin down  (see Heger et al. 2005; Petrovic
2005) to explain the observed rotational
velocities of WDs. While this conclusion has been reached
before (Suijs et al. 2008 ), this is the first time that a rotating stellar
evolutionary model is computed with the inclusion of realistic AGB 
mass-loss rates that account for the modulations associated with the thermal-pulses. We
find that magnetic field induced angular momentum transport is
needed to slow down the CO cores and that the same process is not
capable of spinning up the envelopes. 
Even if we make the exercise to compute what would be the maximum rotational velocity
of the  star in the ideal situation where all the momentum from the core could
be transported out to the envelope at $t=7.80637 \times 10^8$  yr 
($J_{\rm env}^{\rm max} = J_{\rm core} +  J_{\rm env}  = 5.667 \times10^{48}  $ \gcmss)  we
can only increase the rotational velocity from 296 up to 1017 \cms.
We have to conclude that the
remaining momentum in the core is not enough to speed enough the envelope.

We have also carried out a number of ad-hoc experiments where the
magnetic torques are turned-on at the end of the evolution of the star, when the
core is still rotating fast. In none of the three cases considered,
selected to provide the best possible conditions to transfer angular
momentum to the stellar surface, we can get the
envelope rotating. These experiments just strengthen the conclusions
of this paper, we cannot get the envelope rotating and the core
slowed down even under the best case scenario provided by the physics
of the problem.

Note that non-magnetic rotational transport processes have been found
to be negligible for low-mass stars to slow down the stellar core
(Langer et al. 1999; Palacios et al. 2003; 2006). Other mechanisms
besides magnetic torques might exit (see Talon \&  Charbonnel 2005), 
such as angular momentum transfer through internal gravity waves (Zahn et
al. 1997) but will not alleviate the problem we are having to speed up
the envelope, as we have shown above. 

The use of the particular prescription of Vassiliadis and Wood (1993)
for the mass-loss rate is not expected to affect our main
conclusion. First, for the magnetic case the core has already slowed
down  on the RGB so the differences in the evolution of the star
imposed by a different mass-loss prescription will not be noticeable
at all in the rotational speeds of the envelope. Second, although the non-magnetic
case is  more susceptible to be affected by the adopted mass-loss
rates we have shown that this model is not realistic since it leads to the 
formation of a fast rotating CO core which is not supported by observations. 

Note, nonetheless that the 
Vassiliadis and Wood (1993) mass-loss parameterization is the
most widely used for this phase, however, it has its limitations. In particular,
it restrict the maximum amount of mass-loss to the radiation pressure limit  value, 
%$\Mdot = L/cv_{\rm exp}$ 
while it
has been observed that dust driven winds could exceed this limit by a factor of $\sim 2$
(Knapp 1986; Wood et al. 1992; Whitelock, Feast \& Catchpole 1991). 
Larger AGB radius would be expected if higher mass-loss rates
are reached during the thermal-pulses leading to an even
more pronounced decrease of the rotational speeds of the
envelope. Smaller mass-loss rates would have the opposite
effect on the radius but they would delay the evolution of the
star. 

The explicit assumption within this work is that to develop  a
bipolar outflow we require rotational speeds that allow the
formation of a wind-compressed disk  (Bjorkman \& Cassinelli 1993).
However, bipolar flows can be formed, in principle, by other
mechanisms such as the magnetic bomb discussed by Matt et al. (2006)
in which the wind can be effectively collimated by a strong magnetic
field tied to a star eliminating the need of an external disk. In the
Matt et al. (2006) simulations at $t = 0$, 
the core begins rotating at a constant rate of at least 10\% of the
escape speed. Using a representative core mass and radius in our models ($R_{\rm
  core} = 10^9$ cm and $M_{\rm core} = 0.656 \Mo$ ) we
obtain a escape velocity of 4172 \kms which would require in the Matt et
al. (2006) model an initial core rotational velocity of 417 \kms. This core
rotational speed is a factor 2.78 larger than what we obtain in the
best case scenario, the non-magnetic model ($\sim $ 150 \kms), and
a factor 83  larger than that obtained for the magnetic model, $\sim 5$ \kms.  
So even a purely magnetic collimation mechanism for the wind seems to
require core rotational speeds larger than those that a single
rotating star can produce.

A different possibility is that suggested by  Frank (1995) and Soker
(1998) in which the formation of magnetic spots at the surface of AGB stars 
is responsible of the formation of an equatorial density
enhancement. In this scenario, turbulent dynamos of the $\alpha^2
\Omega$ type (Soker \& Zoabi 2002) should generate such magnetic fields since 
the stellar rotation is very small
($\sim 300$ \cms in our case, which is translated into $\sim 10^{-4} \omega_{\rm Kep} $).  
Dust should be efficiently formed above the cold spots giving rise to an 
enhancement on the mass loss rate. However, as discussed by the
authors, these spots have to be formed preferently close to the equator 
and this mechanism would work only for the formation of
elliptical nebulae given that the density ratio between pole a equator is smaller than 2.

Finally, concerning just single stars, the only mechanism available in the literature 
left to produce an equatorial density enhancement in the slow, 
AGB stellar wind, would be the one studied by Matt et al. (2000).
In this study, a non-rotating  AGB star produces a dipole magnetic field that enforces a wind compression
towards the equator. Rotation, convection or a combination of both is necessary to produce
a dynamo able to generate a  magnetic field ($\Omega$, $\alpha$ or $\alpha-\Omega$ dynamos respectively). 
Since we have computed in this paper that rotation is negligible in
AGB stellar envelopes, convection would be the only relevant
mechanism. All AGB stars have convective envelopes, 
independently of its stellar mass.  But, from observations, bipolar PNe
are not formed around all stars. 
If a dipole magnetic field is the mechanism at work for single stars to form
bipolar planetary nebulae we would require the presence of an
extra-parameter in the models to understand better the relatively small
percentage of bipolar PNe formed.

AGB wind asphericities could result as well from the interaction of the AGB star 
with a binary companion (Livio 1993; Soker 1997; Nordhaus \& Blackman 2006; de Marco 2009; 
de Marco et al. 2013). 
The interaction could involve the spin-up of the envelope by tidal forces, Roche lobe overflows, and 
common envelope evolution.
Binary interactions are beyond the scope of the present work and
it will be the subject of a future paper.

In conclusion, according to state-of-the-art stellar evolution calculations of low-intermediate masses 
which includes the effects of rotation and stellar magnetic fields, single stellar rotators cannot be the
precursors of bipolar planetary nebulae under the current MHD models
requiring toroidal fields. If dipolar magnetic fields are invoked, extra-ingredients
are  needed, besides the presence of the magnetic field, to explain
why bipolar PNe are not ubiquitously formed around all stellar masses.

\acknowledgments

G.G.-S. is partially supported by CONACyT grant 178253  and DGAPA grant IN100410.
E.V. work was supported by the Spanish Ministerio de Ciencia e
Innovaci\'on (MICINN), Plan Nacional de Astronom\'{\i}a y Astrof\'{\i}sica, under
grant AYA2010-20630 and by the Marie Curie program under grant FP7-People-RG268111. 
A.M. acknowledge support for this work provided by the Spanish 
Ministry of Economy and Competitiveness under grant AYA-2011-27754. 
We would like to thank our anonymous referee for his valuable comments which improved the
presentation of the paper

\clearpage

\begin{figure}
\epsscale{1.10}
%\plotone{newfig1.eps}
\vspace*{180mm}
\includegraphics{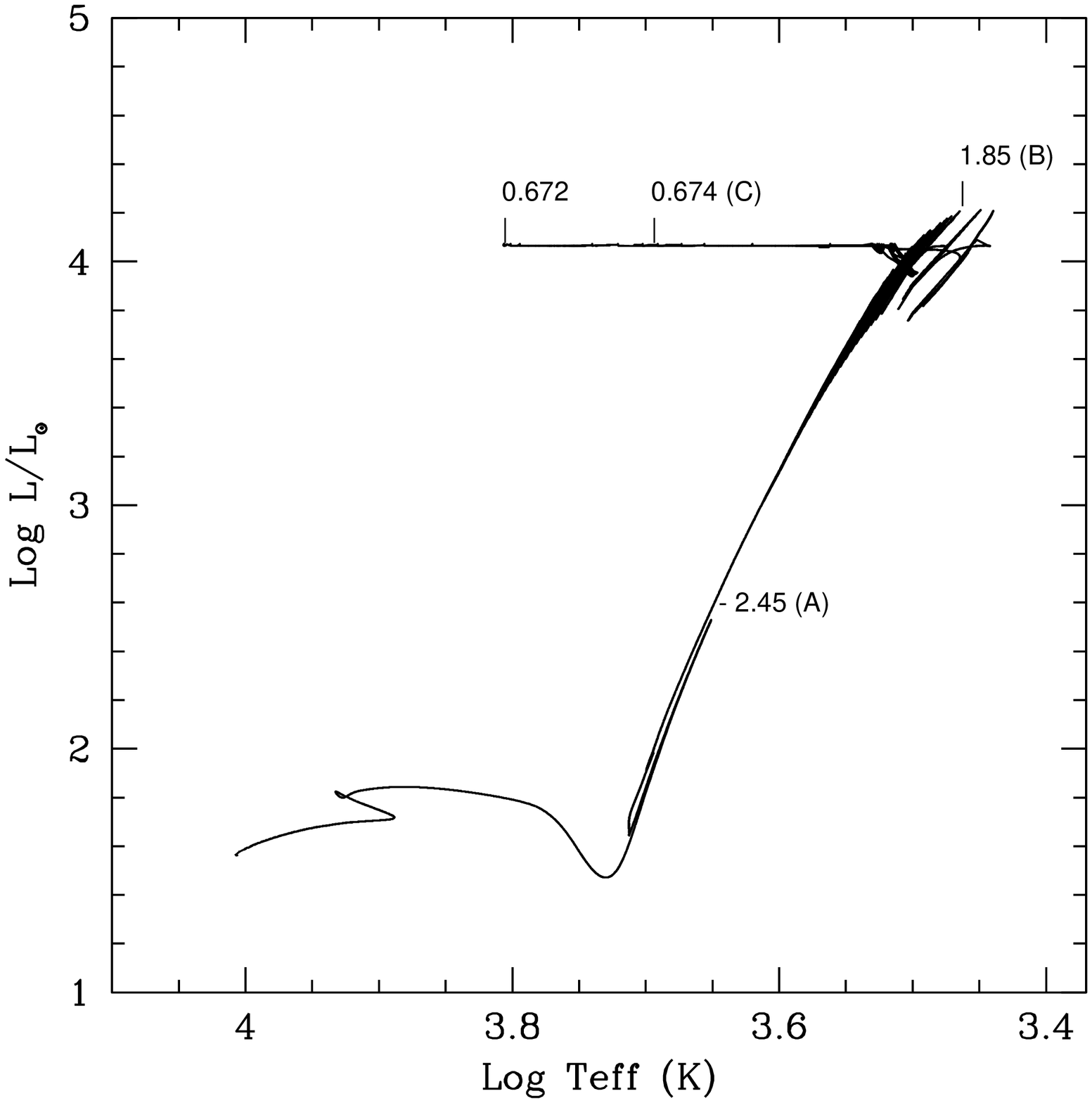}
\caption{Evolutionary track of the magnetic 2.5 \Mo star on
the HR diagram. The ZAMS rotational velocity assumed for this
model is 250 \kms.}
\label{HR}
\end{figure}

\clearpage

\begin{figure}
\epsscale{1.150}
\vspace*{170mm}
%\plotone{newfig2.eps}
\includegraphics{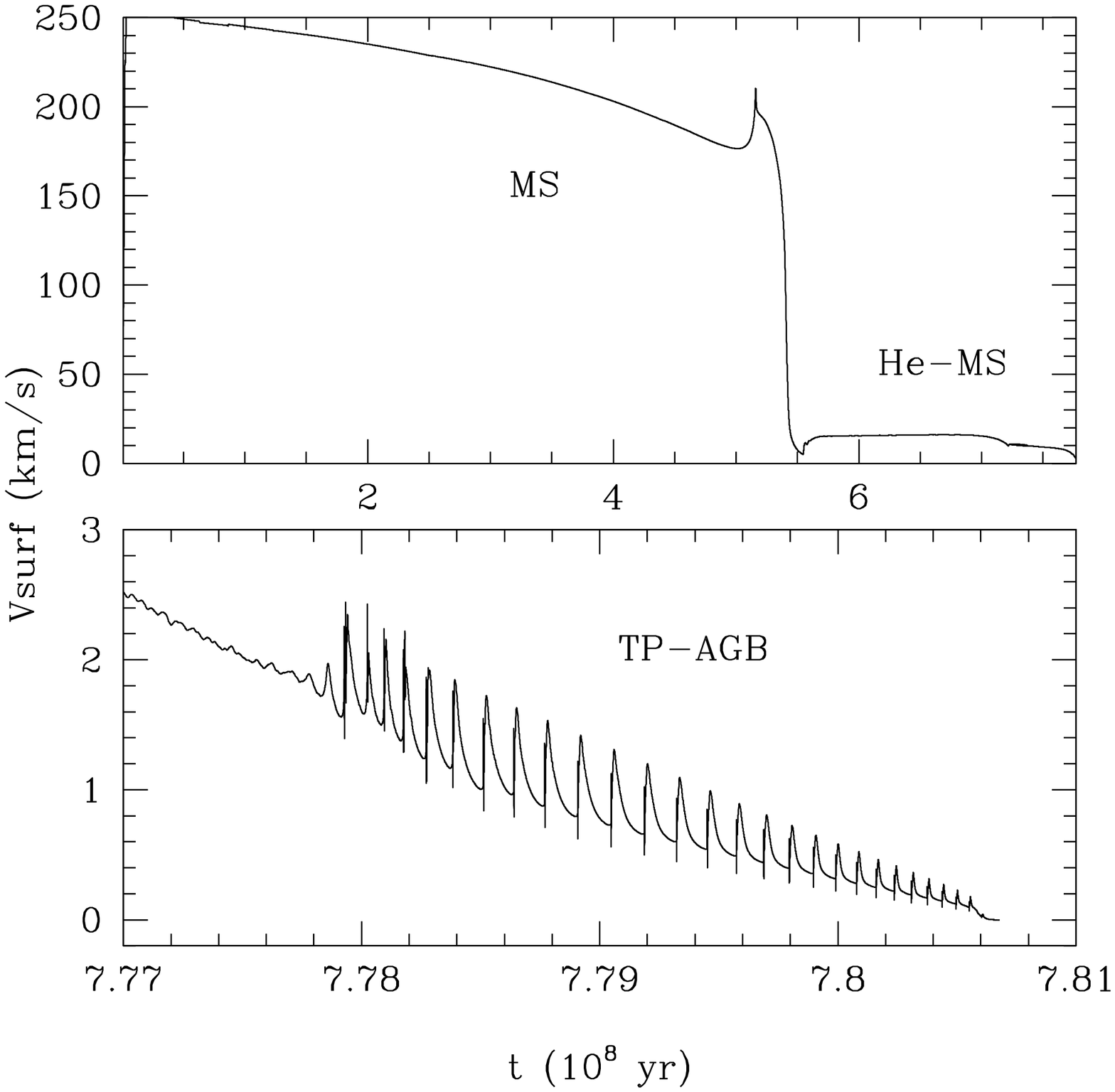}
\caption{\small{Evolution of the surface equatorial rotational velocity
of the 2.5 \Mo model.  The initial equatorial rotational velocity is 250 \kms and
the model does not include magnetic torques. The top panel follows the
evolution to the early TP-AGB phase and the bottom panel shows the
rest of the evolution until the end of the calculation when the star
reaches a temperature of 6400 \,K.}}
\label{vsuf}
\end{figure}

\clearpage

\begin{figure}
\epsscale{1.10}
\vspace*{180mm}
%\plotone{newfig3.eps}
\includegraphics{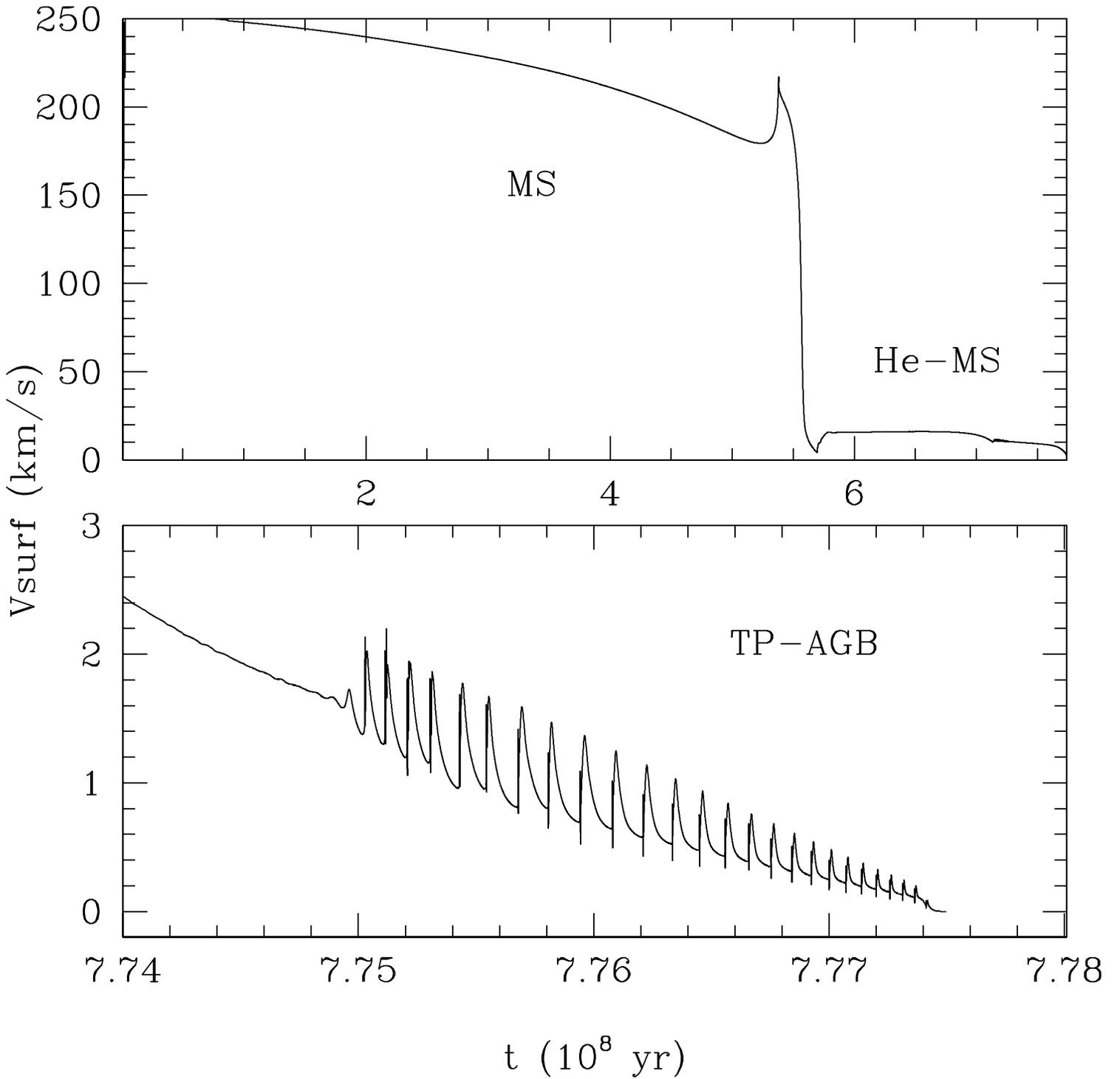}
\caption{Same as Figure 2 but including angular momentum transport induced by 
magnetic torques.}
\end{figure}

\clearpage

\begin{figure}
\epsscale{1.10}
\vspace*{180mm}
%\plotone{newfig4.eps}
\includegraphics{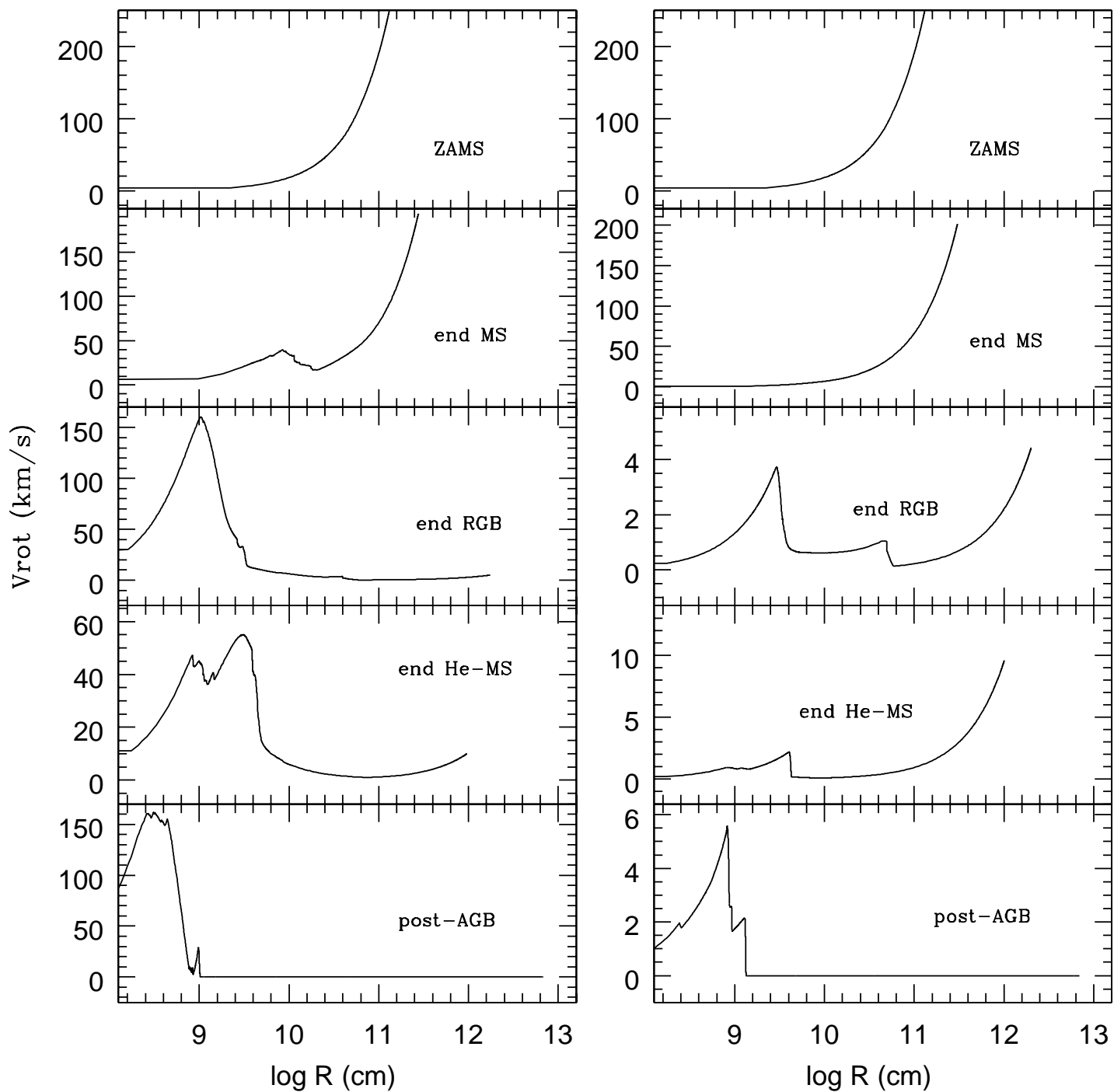}
\caption{Snapshots of the distribution of the rotational velocities
within the stellar structure taken at key evolutionary times.
The non-magnetic case is in the left panels while the magnetic case in the right panels.}
\end{figure}

\clearpage

\begin{figure}
\epsscale{1.10}
\vspace*{180mm}
%\plotone{newfig5.eps}
\includegraphics{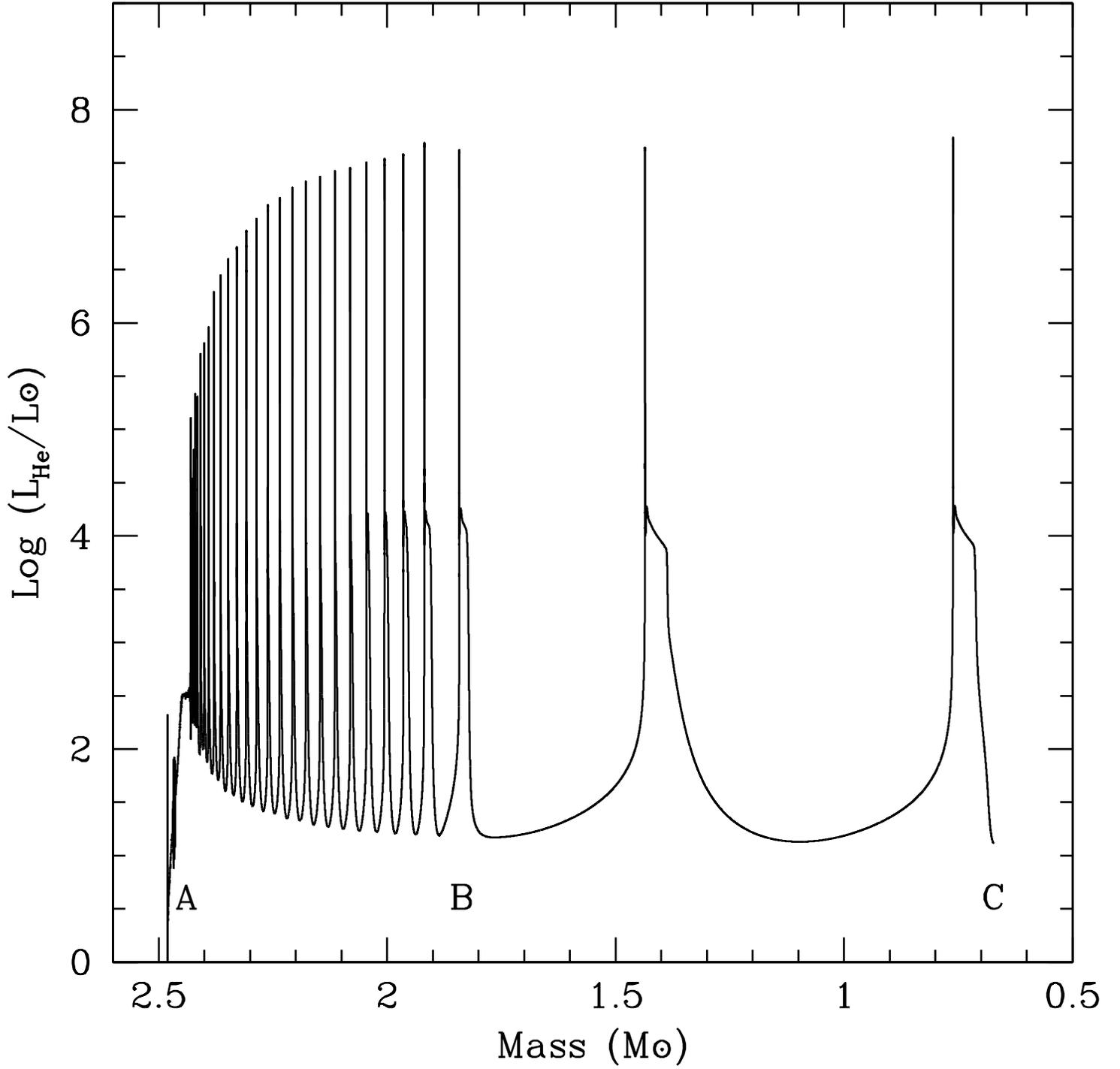}
\caption{Evolution of the Helium burning luminosity as a function of the stellar mass.}
\end{figure}

\clearpage

\begin{figure}
\epsscale{.50}
\vspace*{180mm}
%\plotone{newfig6.eps}
\includegraphics{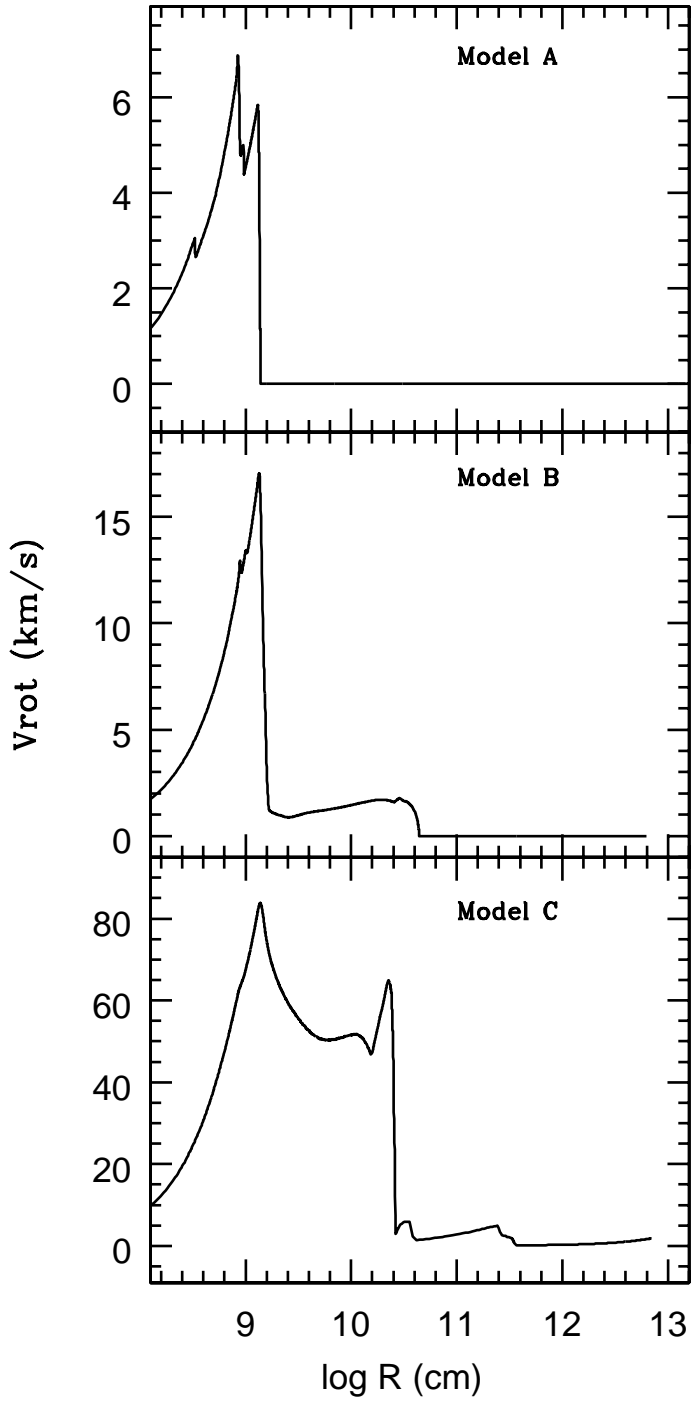}
\caption{The distribution of the rotational velocities
within the stellar structure at the post-AGB phase (6400 \,K).  }
\end{figure}

\clearpage

\begin{figure}
\epsscale{1.150}
\vspace*{170mm}
%\plotone{newfig7.eps}
\includegraphics{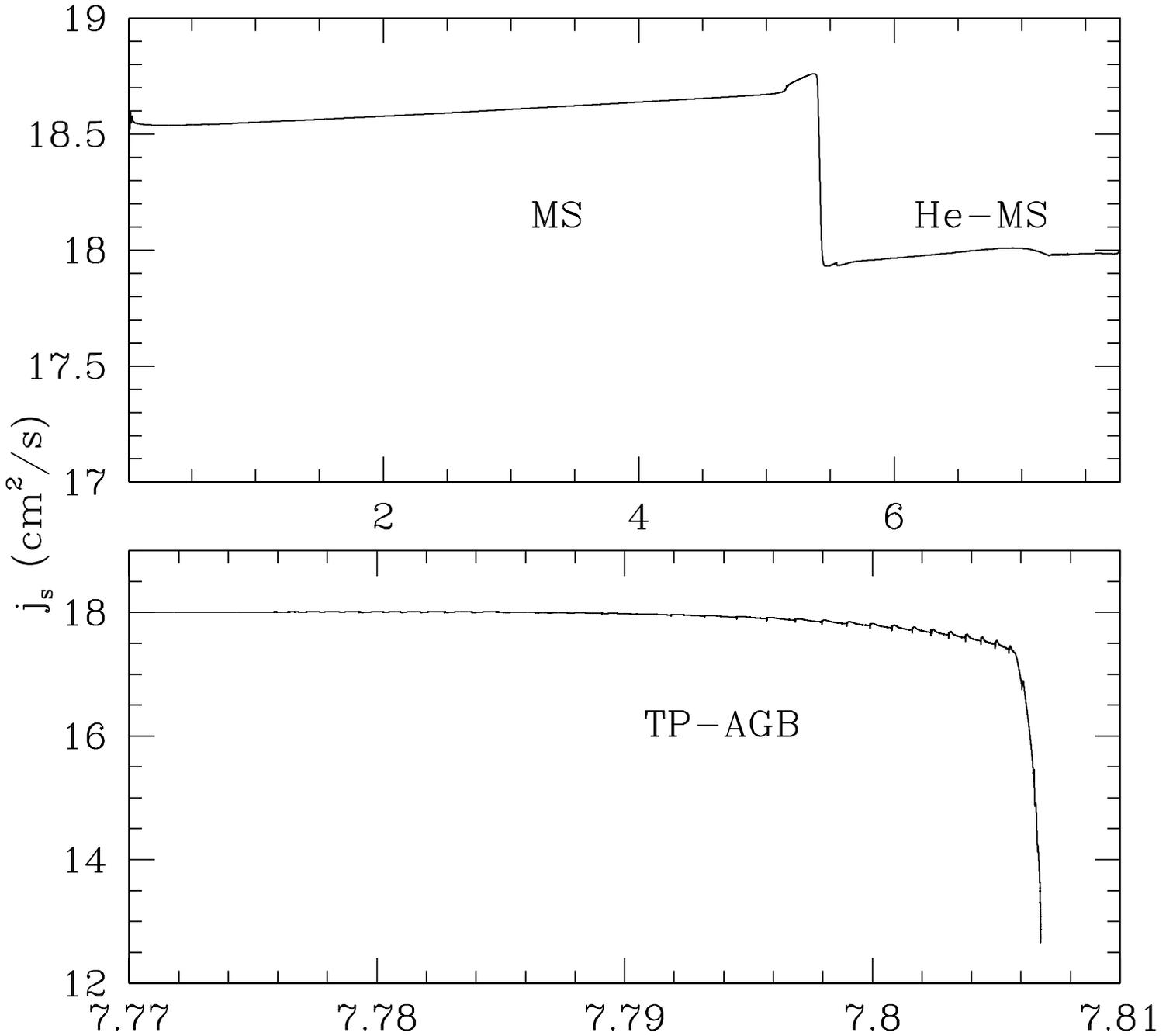}
\caption{\small{Evolution of the specific angular momentum of the surface layers 
for the 2.5 \Mo model without  magnetic torques. The top panel follows the
evolution to the early TP-AGB phase and the bottom panel shows the
rest of the evolution until the end of the calculation when the star
reaches a temperature of 6400 \,K. } }
\end{figure}

\clearpage

\begin{figure}
\epsscale{1.150}
\vspace*{180mm}
%\plotone{newfig8.eps}
\includegraphics{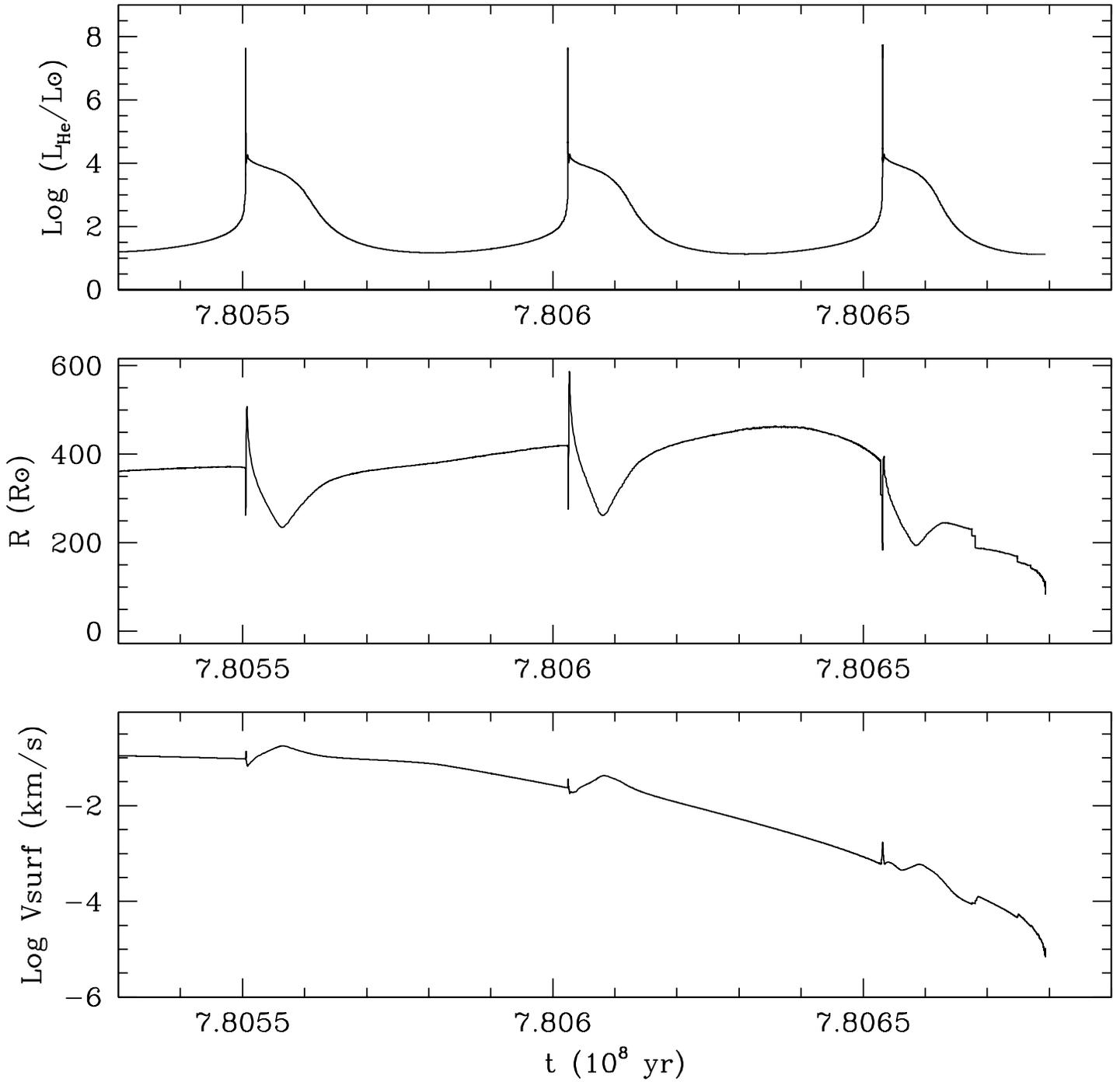}
\caption{Evolution of the stellar radius (middle) and the surface rotational velocity 
(bottom) during the last three thermal pulses, which are shown on the top as the Helium 
luminosity. The last points correspond with the post-AGB at 6400 \,K .} 
\end{figure}

\end{document}